\magnification=1200
\hsize=125mm
\vsize=185mm
\parindent=8mm
\frenchspacing

\font\Bbb=msbm10
\font\toto=cmbx10 scaled 1400
\font\sect=cmbx10 scaled 1200

\def\R{\hbox{\Bbb R}}
\def\S{\hbox{\Bbb S}}
\def\N{\hbox{\Bbb N}}
\def\pa{\partial}
\def\b{\backslash}

\noindent{\toto On inverse problems for the multidimensional}

\noindent{\toto  relativistic Newton equation at fixed energy}
\vskip 7mm

\noindent{Alexandre Jollivet}

\vskip 1cm

\noindent{\bf Abstract.} In this paper, we consider inverse scattering and inverse boundary value problems at sufficiently large and fixed energy for the 
multidimensional relativistic Newton equation with an external potential $V$, $V\in C^2$. Using known results, 
we obtain, in particular, theorems of uniqueness.

\vskip 1cm

\noindent {\bf 1. Introduction}

\vskip 4mm
\noindent 1.1 {\it Relativistic Newton equation.}
Consider the Newton equation in the relativistic case (that is the Newton-Einstein equation) in an open subset $\Omega$ of $\R^n,$
$n\ge 2,$
$$
\eqalign{\dot p =-\nabla V(x),\cr
p={\dot x \over \sqrt{1-{|\dot x|^2 \over c^2}}},\ \dot p={dp\over dt},&\ \dot x={dx\over dt},}
\leqno (1.1)
$$
where
$V\in C^2(\bar \Omega,\R)$ (i.e. there exists $\tilde{V}\in C^2(\R^n,\R)$ such that $\tilde{V}$ restricted to $\bar \Omega$ is equal to $V$) 
and $x=x(t)$ is a $C^1$
function with values in $\Omega.$

By $\|V\|_{C^2}$ we denote the supremum of the set 
$\{|\pa_x^j V(x)|\ |\ x\in \Omega,\ $ $j=(j_1,..,j_n)\in (\N\cup \{0\})^n,\sum_{i=1}^n j_i\le 2\}.$

The equation (1.1) is the equation of motion of a relativistic particle of mass $m=1$
in an external scalar potential $V$ (see [E] and, for example, Section 17 of [LL]). The potential $V$ can be, for example, an electric
potential or a gravitational potential. In this equation
$x$ is the position of the particle, $p$ is its impulse, $t$ is the time and $c$ is the speed of light.

For the equation (1.1) the energy
$$
E=c^2\sqrt{1+{|p(t)|^2 \over c^2}}+V(x(t))
$$
is an integral of motion.
We denote by $B_c$ the euclidean open ball whose radius is c and whose centre is 0.

In this paper we consider the equation (1.1) in two situations.
We study equation (1.1) when
$$
\eqalign{
&\Omega=D {\rm\ where\ } D {\rm\ is\ a\ bounded\ strictly\ convex\ open\ subset\ of\ }\R^n, n\ge 2,\cr
&{\rm with\ }C^2 {\rm\ boundary}.
}\leqno(1.2{\rm a})
$$

And we study equation (1.1) when 
$$
\Omega=\R^n {\rm\ and\ }|\pa_x^jV(x)|\le \beta_{|j|}(1+|x|)^{-\alpha-|j|},\ x\in \R^n,\leqno(1.2{\rm b})
$$
for $|j|\le 2 $ and some $\alpha>1$ (here $j$ is the multiindex $j\in (\N\cup \{0\})^n,$ $|j|=\sum_{i=1}^nj_i$ and $\beta_{|j|}$ are positive real
constants).

For the equation (1.1) under condition (1.2a), we consider boundary data. For
equation (1.1) under condition (1.2b), we consider scattering data.

\vskip 2mm 
\noindent 1.2 {\it Boundary data.}
For the equation (1.1) under condition (1.2a), one can prove that at
sufficiently large energy $E$ (i.e. $E>E(\|V\|_{C^2},D)$), the solutions $x$ of energy
$E$ have the following properties (see Subsections 3.1, 3.2 and 3.3 of Section 3):
$$
\eqalign{
&{\rm for\ each\ solution\ }x(t)\ {\rm there\ are\ }t_1,t_2\in \R, t_1<t_2, {\rm\ such\ that\ }\cr
&x\in C^3([t_1,t_2],\R^n),x(t_1), x(t_2)\in \pa D, x(t)\in D
{\rm \ for\ }t\in ]t_1,t_2[,\cr
&x(s_1)\not=x(s_2) {\rm \ for\ } s_1,s_2\in [t_1,t_2], s_1\not= s_2;
}\leqno(1.3) 
$$
$$
\eqalign{
&{\rm for\ any\ two\ distinct\ points\ }q_0,q\in \bar D, {\rm\ there\ is\ one\ and\ only\ one\ solution}\cr
&x(t)=x(t,E,q_0,q){\rm\ such\ that\ }x(0)=q_0, x(s)=q {\rm\ for\ some\ }s>0.\cr
}\leqno(1.4) 
$$

Let $(q_0,q)$ be two distinct points of $\pa D$. By $s(E,q_0,q)$ we denote the time at which $x(t,E,q_0,q)$ reaches $q$.
By $k(E,q_0,q)$ we denote the velocity vector $\dot x(s(E,q_0,q),E,q_0,q).$ We consider $k(E,q_0,q),$ $q_0,q\in\pa D, q_0\not=q,$ as the boundary value data. 

\vskip 2mm

\noindent 1.3 {\it Scattering data.}
For the equation (1.1) under condition (1.2b), the following is valid (see [Y]): for any 
$(v_-,x_-)\in B_c\times\R^n,\ v_-\neq 0,$
the equation (1.1)  has a unique solution $x\in C^2(\R,\R^n)$ such that
$$
{x(t)=v_-t+x_-+y_-(t),}\leqno (1.5)
$$
where $\dot y_-(t)\to 0,\ y_-(t)\to 0,\ {\rm as}\ t\to -\infty;$  in addition for almost any 
$(v_-,x_-)\in B_c\times \R^n,\ v_-\neq 0,$
$$
{x(t)=v_+t+x_++y_+(t),}\leqno (1.6)
$$
 where $v_+\neq 0,\ |v_+|<c,\ v_+=a(v_-,x_-),\ x_+=b(v_-,x_-),\ \dot y_+(t)\to 0,\ \ y_+(t)\to 0,{\rm\ as\ }t \to +\infty$.

For an energy $E>c^2,$ the map 
$S_E: \S_E\times\R^n \to \S_E\times\R^n $ (where 
$\S_E=\{v\in B_c\ |\ |v|=c\sqrt{1-\left({c^2\over E}\right)^2} \}$)
given by the formulas
$$
{v_+=a(v_-,x_-),\ x_+=b(v_-,x_-)},\leqno(1.7)
$$
is called the scattering map at fixed energy $E$ for the equation (1.1) under condition (1.2b). By ${\cal D}(S_E)$ we denote the domain of definition of $S_E.$  The data
$a(v_-,x_-),$ $b(v_-,x_-)$ for $(v_-,x_-)\in {\cal D}(S_E)$ are called the scattering data at fixed energy $E$ for the equation (1.1) under condition (1.2b).

\vskip 2mm
\noindent 1.4 {\it Inverse scattering and boundary value problems.}
In the present paper, we consider the following inverse boundary value problem at fixed energy for the equation (1.1) under condition (1.2a):
$$
\eqalign{
{\bf Problem\ }1: &{\rm\ given\ }k(E,q_0,q) {\rm\ for\ all\ } (q_0,q)\in\pa D\times \pa D,\cr
& q_0\not=q,{\rm\ at\ fixed
\ sufficiently\ large\ energy\ }E,
{\rm\ find\ } V.
}
$$
The main results of the present work include the following theorem of uniqueness for Problem 1.
\vskip 2mm
{\bf Theorem 1.1.} {\it At fixed} $E>E(\|V\|_{C^2},D)$, {\it the boundary data} $k(E,q,q_0)$, 
$(q_0,q)\in \pa D\times \pa D,$ $q_0\not=q$,
{\it uniquely determine $V$.
}
\vskip 2mm
Theorem 1.1 follows from a reduction of Problem 1 to the problem of determining an isotropic Riemannian metric from its hodograph and from Theorem 3.1 (see Section 3).

In the present paper, we also consider the following inverse scattering problem at fixed energy for the equation (1.1) under condition (1.2b):
$$
{\bf Problem\ }2: {\rm\ given\ }S_E {\rm\ at\ fixed\ energy\ }E, {\rm\ find\ }V.
$$
The main results of the present work include the following theorem of uniqueness for Problem 2.
\vskip 2mm 

{\bf Theorem 1.2.}
{\it Let $\lambda\in \R^+$ and let $D$ be a bounded strictly convex open subset of $\R^n$ with $C^2$ boundary.  Let $V_1, V_2\in C^2_0(\R^n,\R),$ 
$\max(\|V_1\|_{C^2},\|V_2\|_{C^2})\le \lambda,$ and ${\rm supp}(V_1)\cup {\rm supp}
(V_2)\subseteq D.$ Let $S^i_E$ be the scattering map at fixed energy $E$ subordinate to $V_i$ for $i=1,2.$ There exists a nonnegative real
constant $E(\lambda,D)$ such that for any
$E>E(\lambda,D),$ $V_1\equiv V_2$ if and only if $S^1_E\equiv S^2_E.$
}

\vskip 2mm 

Theorem 1.2 follows from Theorem 1.1 and Proposition 2.1.

\vskip 2mm

{\bf Remark 1.1.} Note that for $V\in C^2_0(\R^n,\R)$, if $E< c^2+\sup\{V(x)\ |\ x\in \R^n\}$ then $S_E$ does not determine uniquely $V.$

Note also that reducing Problem 1 to the problem of determining an isotropic Riemannian metric from its hodograph, one
can give also stability estimates for Problem 1 under the assumptions of Theorem 1.1.

\vskip 2mm

\noindent 1.5 {\it Historical remarks.}
An inverse boundary value problem at fixed energy and at high energies was studied in [GN] for the multidimensional nonrelativistic Newton equation in a bounded open
strictly convex domain. In [GN] results of uniqueness and stability for the inverse boundary value problem at fixed energy are derived from results for the
problem of determining an isotropic Riemannian metric from its hodograph (for this geometrical problem, see [MR], [B] and [BG]). 

Novikov [N2] studied inverse scattering for nonrelativistic multidimensional Newton equation. Novikov [N2] gave, in particular, a connection between the inverse scattering
problem at fixed energy and Gerver-Nadirashvili's inverse boundary value problem at fixed energy. Theorem 1.2 of the present work is a generalization of theorem 5.2 of [N2] to the
relativistic case.

Inverse scattering at high energies for the relativistic multidimensional Newton equation was studied by the author (see [J1], [J2]).

As regards analogs of Theorems 1.1, 1.2 and Proposition 2.1 for nonrelativistic quantum mechanics see [N1], [NSU], [N3] and further
references therein. As regards an analog of Theorem 1.2 for relativistic quantum mechanics see [I]. As regards results given in the
literature on inverse scattering in quantum mechanics at high energy limit see references given in [J2].
\vskip 2mm

\noindent 1.6 {\it Structure of the paper.}
The paper is organized as follows. In Section 2, we give some properties of boundary data and scattering data
and we connect the inverse scattering problem at fixed energy to the inverse boundary value problem at fixed energy.
In Section 3, we obtain a theorem of uniqueness and stability (Theorem 3.1) for the inverse boundary value problem. 
This theorem is a generalization to relativistic case of theorem 4 of [GN]. 
\vskip 4mm

\noindent{\bf Acknowledgement.} This work was fulfilled in the framework of Ph. D. thesis researchs under the direction of R.G. Novikov.

\vskip 1cm

\noindent{\bf 2. Scattering data and boundary value data.}

\vskip 4mm
\noindent 2.1 {\it Properties of the boundary value data.} Let $D$ be a bounded strictly convex open subset of $\R^n$, 
$n\ge 2$, with $C^2$ boundary.

At fixed sufficiently large $E$ (i.e. $E>E(\|V\|_{C^2},D)$ $\ge c^2+\sup_{x\in \bar D}V(x)$) solutions $x(t)$
of the equation (1.1) under condition (1.2a) have the following properties (see Subsections 3.1, 3.2 and 3.3 of Section 3):

$$
\eqalign{
&{\rm for\ each\ solution\ }x(t)\ {\rm there\ are\ }t_1,t_2\in \R, t_1<t_2, {\rm\ such\ that\ }\cr
&x\in C^3([t_1,t_2],\R^n),x(t_1), x(t_2)\in \pa D, x(t)\in D
{\rm \ for\ }t\in ]t_1,t_2[,\cr
&x(s_1)\not=x(s_2) {\rm \ for\ } s_1,s_2\in [t_1,t_2], s_1\not= s_2,\dot x(t_1)N(x(t_1))<0\cr
&{\rm and\ }\dot x(t_2)N(x(t_2))>0,{\rm\ where\ }N(x(t_i)) {\rm\ is\ the\ unit\ outward}\cr
&{\rm normal\ vector\ of \ }\pa D{\rm\ at\ }x(t_i) {\rm\ for\ }i=1,2;
}\leqno(2.1) 
$$
$$
\eqalign{
&{\rm for\ any\ two\ points\ }q_0,q\in \bar D, q\not=q_0, {\rm\ there\ is\ one\ and\ only\ one\ solution}\cr
&x(t)=x(t,E,q_0,q){\rm\ such\ that\ }x(0)=q_0, x(s)=q {\rm\ for\ some\ }s>0;\cr
&\dot x(0,E,q_0,q)\in C^1((\bar D\times\bar D)\backslash \bar G,\R^n),{\rm\ where\ } \bar G {\rm\ is\ the\ diagonal\ in\ }\bar D\times\bar D,
}\leqno(2.2) 
$$
(where by ``$\dot x(0,E,q_0,q)\in C^1((\bar D\times\bar D)\backslash \bar G,\R^n)$" we mean that there exists an open neighborhood
$\Omega$ of $\bar D$
such that $\dot x(0,E,q_0,q)$ is the restriction to $(\bar D\times\bar D)\backslash \bar G$ of a function which belongs to 
$C^1((\Omega\times \Omega)\backslash\Delta)$ where $\Delta$ is the diagonal of $\Omega\times \Omega$).
Let $E> E(\|V\|_{C^2},D)$. Consider the solution $x(t,E,q_0,q)$ from (2.2) for $q_0,q\in \pa D,$ $q_0\not=q.$ We remind that
$s=s(E,q_0,q)$ is the root of the
equation 
$$
x(s,E,q_0,q)=q,\ \ s>0,
$$ 
and we remind that $k(E,q_0,q)=\dot x(s(E,q_0,q),E,q_0,q).$ We consider $k(E,q_0,q),$ $q_0,q\in \pa D,$ $q_0\not=q$ as the boundary value data.

Let $k_0(E,q_0,q)=\dot x(0,E,q_0,q).$ 
Note that 
$$
\eqalign{
k_0(E,q_0,q)=&-k(E,q,q_0),\cr
|k_0(E,q_0,q)|=&c\sqrt{1-\left({E-V(q_0)\over c^2}\right)^{-2}},
}\leqno(2.3)
$$
for $E>E(\|V\|_{C^2},D)$ and 
$(q,q_0)\in (\pa D\times \pa D)\backslash \pa G.$

\vskip 4mm

\noindent 2.2 {\it Boundary data for the non relativistic case.}
If one considers the nonrelativistic Newton equation in $D$ instead of the equation (1.1) under condition (1.2a), one obtains the existence of a constant
$E'(\|V\|_{C^2},D)$ such that the solutions $x(t)$ of the nonrelativistic Newton equation with energy $E={1\over 2}|\dot x(t)|^2+V(x(t))$, $E>E'(\|V\|_{C^2},D)$, 
also have properties (2.1) and (2.2) (see [GN]). Hence one can define the time  $s'(E,q_0,q)$ and the vector $k'(E,q_0,q)$ for
$E>E'(\|V\|_{C^2},D),$ $(q_0,q)\in (\pa D\times \pa D)\backslash \pa G,$ as were defined $s(E,q_0,q)$, $k(E,q_0,q)$ for
$E>E(\|V\|_{C^2},D),$ $(q_0,q)\in (\pa D\times \pa D)\backslash \pa G.$ In [GN], $s'(E,q_0,q)$ and $k'(E,q_0,q)$ for
$E>E'(\|V\|_{C^2},$ $D),$ $(q_0,q)\in (\pa D\times \pa D)\backslash \pa G,$ were taken as boundary value data for the
multidimensional nonrelativistic Newton equation and [GN] obtains, in particular, that $s'(E,q_0,q)$ given for all 
$E>E'(\|V\|_{C^2},D),$ $(q_0,q)\in (\pa D\times \pa D)\backslash \pa G$ uniquely determines $V$ and that
$k'(E,q_0,q)$ given for all $(q_0,q)\in (\pa D\times \pa D)\backslash \pa G$ uniquely determines $V$ on $\bar D$ at fixed energy
$E>E'(\|V\|_{C^2},D).$

\vskip 4mm

\noindent 2.3 {\it Properties of the scattering operator.}
For equation (1.1) under condition (1.2b), the following is valid (see [Y]): for any 
$(v_-,x_-)\in B_c\times\R^{d},\ v_-\neq 0,$
the equation (1.1) under condition (1.2b) has a unique solution $x\in C^2(\R,\R^d)$ such that
$$
{x(t)=v_-t+x_-+y_-(t),}\leqno (2.4)
$$
where $\dot y_-(t)\to 0,\ y_-(t)\to 0,\ {\rm as}\ t\to -\infty;$  in addition for almost any 
$(v_-,x_-)\in B_c\times \R^{d},\ v_-\neq 0,$
$$
{x(t)=v_+t+x_++y_+(t),}\leqno (2.5)
$$
where $v_+\neq 0,\ |v_+|<c,\ v_+=a(v_-,x_-),\ x_+=b(v_-,x_-),\ \dot y_+(t)\to 0,\ \ y_+(t)\to 0,{\rm\ as\ }t \to +\infty$.

The map $S: B_c\times\R^d \to B_c\times\R^d $ given by the formulas
$$
{v_+=a(v_-,x_-),\ x_+=b(v_-,x_-)}\leqno(2.6)
$$
is called the scattering map for the equation (1.1) under condition (1.2b). The functions $a(v_-,x_-),$ 
$b(v_-,x_-)$ are called the scattering data for the equation (1.1) under condition (1.2b).

By ${\cal D}(S)$ we denote the domain of definition of $S$; by ${\cal R}(S)$ we denote the range of $S$ (by definition,
if $(v_-,x_-)\in {\cal D}(S)$, then $v_-\neq 0$ and $a(v_-,x_-)\neq 0$). 

The map $S$  has the following simple properties (see [Y]):
for any $(v,x)\in B_c\times \R^d$, $(v,x)\in {\cal D}(S)$ if and only if $(-v,x)\in {\cal R}(S)$;
${\cal D}(S)$  is an open set of $B_c \times \R^d$ and ${\rm Mes}((B_c \times \R^d) \b {\cal D}(S))=0$ for the Lebesgue
measure on $B_c \times \R^d$  induced by the Lebesgue measure on $\R^d\times\R^d$;
the map $S:{\cal D}(S)\to {\cal R}(S)$ is continuous and preserves the element of volume; for any $(v,x)\in {\cal D}(S)$,
$a(v,x)^2=v^2.$

The map $S$ restricted to 
$$
\Sigma_E=\{(v_-,x_-)\in B_c\times \R^d\ |\ |v_-|=c\sqrt{1-\left({c^2\over E}\right)^2}\}
$$
is the scattering operator at fixed energy $E$ and is denoted by $S_E$.

We will use the fact that the map $S$ is uniquely determined by its restriction to ${\cal M}(S)={\cal
D}(S)\cap {\cal M},$ where
$$
{\cal M}=\{(v_-,x_-)\in B_c \times \R^d|v_-\neq 0, v_-x_-=0\}.
$$
This observation is completely similar to the related observation of [N2], [J1] and is based on the fact that if $x(t)$ satisfies (1.1), then $x(t+t_0)$
also satisfies (1.1) for any $t_0\in\R$.
In particular, the map S at fixed energy $E$ is uniquely determined by its restriction to ${\cal M}_E(S)={\cal D}(S)\cap {\cal M}_E,$
where ${\cal M}_E=\Sigma_E\cap {\cal M}.$

\vskip 4mm

\noindent 2.4 {\it Inverse scattering problem and inverse boundary value problem.}
Assume that 
$$
V\in C^2_0(\bar D,\R).\leqno(2.7)
$$
We consider equation (1.1) under condition (1.2a) and equation (1.1) under condition (1.2b). We shall connect  the boundary value data $k(E,q,q_0)$ for $E>E(\|V\|_{C^2},D)$ and 
$(q,q_0)\in (\pa D\times \pa D)\backslash \pa G,$ to the scattering data $a,b.$

\vskip 2mm

{\bf Proposition 2.1.} 
{\it Let $E>E(\|V\|_{C^2},D).$ Under condition} (2.7), {\it the following statement is valid:
$s(E,q_0,q),$ $k(E,q_0,q)$ given for all $(q,q_0)\in (\pa D\times \pa D)\backslash \pa G,$ are determined uniquely by the scattering data 
$a(v_-,x_-)$, $b(v_-,x_-)$ given for all $(v_-,x_-)\in {\cal M}_E(S).$
The converse statement holds:  $s(E,q_0,q),$ $k(E,q_0,q)$ given for all $(q,q_0)\in (\pa D\times \pa D)\backslash \pa G,$ determine uniquely the scattering data 
$a(v_-,x_-)$, $b(v_-,x_-)$ for all $(v_-,x_-)\in {\cal M}_E(S).$
}

\vskip 2mm

{\it Proof of Proposition} 2.1.
First of all we introduce functions $\chi,$ $\tau_-$ and $\tau_+$ dependent on $D.$

For $(v,x)\in \R^n\backslash \{0\}\times\R^n$, $\chi(v,x)$ denotes the nonnegative number of points contained in the intersection of
$\pa D$ with the straight line parametrized by $\R\to \R^n, t\mapsto tv+x.$ As $D$ is a strictly convex open subset of $\R^n$ with $C^2$ boundary, $\chi(v,x)\le 2$ 
for all $v,x\in \R^n,$ $v\not=0.$

Let $(v,x)\in \R^n\backslash \{0\}\times\R^n$. Assume that $\chi(v,x)\ge 1$. The real $\tau_-(v,x)$ denotes the smallest real number $t$ such
that $\tau_-(v,x)v+x\in \pa D$, and the real $\tau_+(v,x)$ denotes the greatest real number $t$ such
that $\tau_+(v,x)v+x\in \pa D$ (if $\chi(v,x)=1$ then $\tau_-(v,x)=\tau_+(v,x)$).

\noindent {\it Direct statement.} Let $(q_0,q)\in (\pa D\times \pa D)\backslash \pa G.$
Under conditions (2.7) and from (2.1) and (2.2), it follows that there exists a unique $(v_-,x_-)\in {\cal M}_E(S)$ such that
$$
\eqalign{
&\chi(v_-,x_-)=2,\cr
&q_0=x_-+\tau_-(v_-,x_-)v_-,\cr
&q=b(v_-,x_-)+\tau_+(a(v_-,x_-), b(v_-,x_-))a(v_-,x_-).\cr
}
$$
In addition, $s(E,q_0,q)=\tau_+(v_-,x_-)-\tau_-(v_-,x_-)$ and $k(E,q_0,q)=a(v_-,x_-).$

\noindent {\it Converse statement.} Let $(v_-,x_-)\in {\cal M}_E(S).$ 
Under conditions (2.7), if $\chi(v_-,x_-)$ $\le 1$ then $(a(p_-,x_-),b(p_-,x_-))=(p_-,x_-).$

Assume that $\chi(v_-,x_-)=2.$ Let 
$$
q_0=x_-+\tau_-(v_-,x_-)v_-.
$$
From (2.1) and (2.2) it follows that there is one and only one solution  of the equation
$$
-k(E,q,q_0)=v_-,\ q\in \pa D,\ q\not=q_0.\leqno(2.8)
$$
We denote by $q(v_-,x_-)$ the unique solution of (2.8).
Hence we obtain 
$$
\eqalign{
a(v_-,x_-)=&k(E,q_0,q(v_-,x_-)),\cr
b(v_-,x_-)=&q(v_-,x_-)-k(E,q_0,q(v_-,x_-))\cr
&\times(s(E,q_0,q(v_-,x_-))+\tau_-(v_-,x_-)).\cr
}
$$
Proposition 2.1 is proved. \hfill $\sqcap\hskip -2.32mm\sqcup$

\vskip 2mm

For a more complete discussion about connection between scattering data and boundary value data, see [N2] considering the non relativistic Newton equation.

\vskip 1cm

\noindent {\bf 3. Inverse boundary value problem.}

\vskip 4mm
\noindent In this Section, Problem 1 of Introduction is studied. Following [GN], we reduce the inverse boundary value
problem to the problem of determining an isotropic Riemannian metric from its hodograph.

\vskip 4mm
\noindent 3.1 {\it Hamiltonian system.} 
Let $E>c^2+\sup_{x\in \bar D}V(x).$ Take $\tilde{V}\in C^2(\R^n,\R)$ such that $\tilde{V}_{|\bar D}\equiv V.$ We shall still denote
$\tilde{V}$ by $V$. Take an open neighborhood $\Omega$ of $\bar D$ such
that $E>c^2+\sup_{x\in \Omega}V(x).$ The equation (1.1) in $\Omega$ is the Euler-Lagrange equation for the Lagrangian $L$ defined by
$L(\dot x,x)=-c^2\sqrt{1-{\dot x^2\over
c^2}}-V(x),$ $\dot x\in B_c$ and $x\in \Omega.$ The Hamiltonian $H$ associated to the Lagrangian $L$ by Legendre's transform (with
respect to $\dot x$) is $H(p,x)=c^2\sqrt{1+{p^2\over c^2}}+V(x)$ where $p\in \R^n$ and $x\in \Omega.$
Then equation (1.1) in $\Omega$ is equivalent to the Hamilton's equation
$$
\eqalign{
\dot x=&{\pa H\over \pa p}(p,x),\cr
\dot p=&-{\pa H\over \pa x}(p,x).
}\leqno(3.1)
$$

\vskip 2mm
\noindent 3.2 {\it Maupertuis's principle.}
In this subsection we apply the Maupertuis's principle to the Hamiltonian system (3.1).

Let $(p(t),x(t)),$ $t\in[t_1,t_2],$ be a solution of (3.1). Let $\gamma(t)=(p(t),x(t),t),$  $t\in[t_1,t_2].$ Then 
$$
\eqalign{
&\gamma {\rm\ is\ a\ critical\ point\ of\ the\ functional\ } J {\rm\ defined\ by\ }\cr
&J(\gamma')=\int_{\gamma'}pdx-H(p,x)dt {\rm\ on\ the\ set\ of\ the\ }C^1 {\rm\ functions\ }\cr
&\gamma':[t_1,t_2]\to \R\times
\Omega\times[t_1,t_2],\ t\mapsto (p'(t),x'(t),t)\cr
& {\rm with\ boundary\ conditions\ }x'(t_1)=x(t_1) {\rm\ and\ } x'(t_2)=x(t_2). 
}\leqno(3.2)
$$
Let $\Sigma$ denote the $2n-1$-dimensional smooth manifold $\{(p,x)\in \R^n\times \Omega\ |\ H(p,x)=E\}$.
From (3.2), it follows that  
$$
\eqalign{
&{\rm for\ any\  }(p(t),x(t)),\ t\in[t_1,t_2], {\rm\  solution\ of\ (3.1)\ with\ energy\ }E\cr
&{\rm and\ for\ any\ strictly\ increasing\ }C^1{\rm\ function\ }\phi {\rm\ from\ some\  closed\ interval}\cr
&[t_-,t_+] {\rm\ of\ }\R {\rm\ onto\ } [t_1,t_2], {\rm\ the\ }C^1 {\rm\ map\ } \bar \gamma {\rm\ defined\ by\ }\cr
&\bar \gamma(t)=(p(\phi(t)),x(\phi(t))), t\in[t_-,t_+], {\rm\ is\ a\ critical\ point\ for\ the\ functional\ }\cr
&\bar J {\rm\ defined\ by\ }\bar J(\gamma')=\int_{\gamma'}pdx {\rm\ on\ the\ set\ of\ the\ }C^1 {\rm\ functions\ }\cr
&\gamma':[t_-,t_+]\to \Sigma,\ t\mapsto (p'(t),x'(t)) {\rm\ with\ boundary\ conditions\ }\cr
&x'(t_-)=x(t_1) {\rm\ and\ } x'(t_+)=x(t_2). 
}\leqno(3.3)
$$
Let $y\in C^2([t_1,t_2],\Omega)$ be such that $\dot y(t)\not=0,$ $t\in [t_1,t_2].$ Let $\phi_y$ be the strictly increasing $C^1$ function
from $[t_1,t_+]$ ($t_+>0$) onto $[t_1,t_2]$ defined by $\phi_y(t_1)=t_1$ and 
$H({\pa L\over \pa\dot x}(\dot \phi(t)\dot y(\phi(t)), y(\phi(t))),y(\phi(t)))=E,$ $t\in[t_1,t_+]$, i.e. $\phi_y$ is the function which
satisfies the ordinary differential equation $\dot \phi_y(t)=c^2{\sqrt{1-({E-V(y(\phi_y(t)))\over c^2})^{-2}}\over|\dot y(\phi_y(t))|},$
$t\in[t_1,t_+]$, with initial datum $\phi_y(t_1)=t_1$.  Let $\bar \gamma(t)=({\pa L\over\pa \dot x}(\dot \phi(\phi^{-1}(t))\dot
y(t),y(t)),$ $y(t)),$ $t\in [t_1,t_2]$.
Then, $\bar J(\bar \gamma(t))=\int_{t_1}^{t_+}{|\dot\phi(t)\dot y(\phi(t))|\over \sqrt{1-{\dot y(\phi(t))^2\dot\phi(t)^2\over c^2}}}
|\dot\phi(t)\dot y(\phi(t))|dt.$ Hence, using that $H(\bar \gamma(t))=E$, $t\in[t_1,t_2]$, we obtain that
$$
\leqalignno{
\bar J(\bar \gamma(t))=&\int_{t_1}^{t_+}r_{V,E}(y(\phi(t))|\dot y(\phi(t))|\dot \phi(t)dt\cr
=&\int_{t_1}^{t_2}r_{V,E}(y(t))|\dot y(t)|dt&(3.4)
}
$$
where $r_{V,E}(x)=c\sqrt{\left({E-V(x)\over c^2}\right)^2-1},$ $x\in \Omega.$

From (3.3) and (3.4), it follows that if $x(t)$, $t\in [t_1,t_2]$, is a solution of (1.1) in $\Omega$ with energy $E$, then $x(t)$ is a
critical point of the functional $l(y)=\int_{t_1}^{t_2}r_{V,E}(y(t))|\dot y(t)|dt$ defined on the set of the functions
$y\in C^1([t_1,t_2],\Omega)$ with boundary conditions $y(t_1)=x(t_1)$ and $y(t_2)=x(t_2)$ (Maupertuis's principle).
As $l(y)$ is the Riemannian length of the curve parametrized by $y\in C^1([t_1,t_2],\Omega)$ for the Riemannian metric $r_{V,E}(x)|dx|$
in $\Omega$, one obtains that if  $x(t)$, 
$t\in [t_1,t_2]$, is a solution of (1.1) with energy $E$, 
then  $x(t)$ composed with its parametrization by arclength (for the Riemannian metric $r_{V,E}(x)|dx|$ in $\Omega$) 
gives a geodesic of the Riemannian metric $r_{V,E}(x)|dx|$ in $\Omega$.

For any solution $x:[0,t_+]\to \Omega$ of equation (1.1) in $\Omega$ with energy $E$, the parametrization by arclength of $x(t)$ is
given by the strictly increasing $C^2$ function $\psi_x$ from $[0,t_+']$ 
($t_+'>0$) onto $[0,t_+]$  defined by the ordinary differential equation
$\dot \psi_x(t)={{E-V(x(\psi_x(t))\over c^4}\over \left({E-V(x(\psi_x(t))\over c^2}\right)^2-1}$ with initial datum $\psi_x(0)=0.$

Applying Maupertuis's principle we obtained the following Lemma.
\vskip 2mm

{\bf Lemma 3.1.}{\it\ Under the assumption $E>c^2+\sup_{x\in \Omega}V(x)$ the following statement is valid:
for any solution $x:[0,t_+]\to \Omega$ of equation} (1.1) {\it in $\Omega$ with energy $E$, the map $y:[0,t_+']\to \Omega$ defined by 
$y(t)=x(\psi_x(t)),$ $t\in[0,t_+'],$ is a geodesic of the Riemannian metric 
$r_{V,E}(y)|dy|$ in $\Omega$ which satisfies 
$r_{V,E}(y)|\dot y|\equiv 1$.
}

\vskip 2mm
We obtain, in particular, that trajectories $\{x(t)\}$ of the 
multidimensional relativistic Newton equation in $\Omega$  with energy $E$ coincide with the geodesics of Riemannian metric 
$r_{V,E}(x)|dx|$ in $\Omega$ where $|dx|$ is the 
canonical euclidean metric on $\Omega$.
(In connection with the Maupertuis's principle and analog of Lemma 3.1 for the Newton equation in the nonrelativistic case, see for example Section 45 of [A].)
\vskip 2mm
\noindent 3.3 {\it Simple metrics.}
We recall the definition of a simple metric $g$ in a bounded open subset $U$ of $\R^n$ with $C^2$ boundary (denoted by $\pa U$) (see
for example [SU]).

Let $U$ be a bounded open subset of $\R^n$ with $C^2$ boundary (denoted by $\pa U$) and let $g$ be a $C^2$ Riemannian metric in $\bar U$.
For $x\in \pa U$ the second fundamental form  $\Pi$ (with respect to $g$) of the boundary at $x$ is defined on the tangent space $T_x(\pa U)$ of $\pa U$ at $x$ by the 
formula
$$
\Pi(\zeta)=g_{x}(\nabla_\zeta N(x),\zeta)
$$
where $\zeta\in T_x(\pa U)$ and $N(x)$ denotes the unit outward normal vector  to the
boundary at $x$ ($g_{x}(N(x),$ $N(x))=1$), and where $\nabla N$ denotes the covariant derivative of the vector field $N$ with respect to the Levi-Civita
connection of the metric $g$. 

We say that $g$ is {\it simple} in $\bar U$, if the second fundamental form is positive definite at every point
$x\in \pa U$ and every two points $x,$ $y\in \bar U$ are joint by an unique geodesic smoothly depending on $x$ and $y$. The latter means
that the mapping  $exp_x:exp_x^{-1}(\bar U)\subseteq T_x\bar U\to \bar U$ is a diffeomorphism for any $x\in \bar U$, where $exp_x(v)$ denotes the point
which is reached at time $1$ by the geodesic in $\bar U$ which starts at $x$ with the velocity $v$ at time $0$ ($T_x\bar U$ denotes the
tangent space of $\bar U$ at the point $x$).

As it was mentioned in [SU], if a Riemannian metric $g$ is close enough to a fixed simple metric $g_0$ in $C^2(\bar U)$, then $g$ is  also
simple.

Here, as $D$ is assumed to be a bounded strictly convex open subset of $\R^n$
with $C^2$ boundary, it follows that the euclidean metric $|dx|$ is simple in $\bar D$.
Hence, from the fact mentioned in [SU], it follows that there exists $E(\|V\|_{C^2},D)$ such that for $E>E(\|V\|_{C^2},D)$ the metric
${cr_{V,E}(x)\over E}|dx|=c^2\sqrt{\left({1-{V(x)\over E}\over c^2}\right)^2-{1\over E^2}}|dx|$ is also simple in $\bar D$.

Hence for $E>E(\|V\|_{C^2},D)$ the metric
$r_{V,E}(x)|dx|$ is simple in $\bar D$. 

Then one can consider properties (2.1) and (2.2) as consequences of Lemma 3.1 and the fact that the metric $r_{V,E}(x)|dx|$ is simple in $\bar D$.

Let $l_{V,E}$ denote the distance on $\bar D$ induced by the Riemannian metric 

\noindent $r_{V,E}(x)|dx|.$

\vskip 2mm

\noindent 3.4 {\it Properties of $l_{V,E}$ at fixed and sufficiently large energy $E$.}
Let 
$E>E(\|V\|_{C^2},$ $D).$ From properties (2.1) and (2.2) (or from the fact that $r_{V,E}(x)|dx|$ is simple), it follows that
$$
\leqalignno{
&l_{V,E}\in C(\bar D\times \bar D,\R),&(3.5)\cr
&l_{V,E}\in C^2((\bar D\times \bar D)\backslash \bar G,\R),&(3.6)\cr
&\max(|{\pa l_{V,E}\over \pa x_i}(\zeta,x)|,|{\pa l_{V,E}\over \pa \zeta_i}(\zeta,x)|)\le C_1,&(3.7)\cr
&|{\pa^2 l_{V,E}\over \pa \zeta_i\pa x_j}(\zeta,x)|\le {C_2\over |\zeta -x|},&(3.8)\cr
}
$$
for $(\zeta,x)\in (\bar D\times \bar D)\backslash \bar G,$ $\zeta=(\zeta_1,..,\zeta_d),$ $x=(x_1,..,x_d),$ and $i=1..n,$ $j=1..n,$ and where $C_1$ and $C_2$ depend on $V$ and $D;$ 
the map $\nu_{V,E}:\pa D\times D\to \S^{n-1},$  defined by
$$
\nu_{V,E}(\zeta,x)={-1\over r_{V,E}(x)}({\pa l_{V,E}\over \pa x_1}(\zeta,x),..,{\pa l_{V,E}\over \pa x_n}(\zeta,x))\leqno(3.9)
$$
has the following properties:
$$
\leqalignno{
&\nu_{V,E}\in C^1(\pa D\times D,\S^{n-1}),&(3.10{\rm a})\cr
}
$$ 
$$
\eqalign{
& {\rm the\ map\ } \nu_{V,E,x}:\pa D\to \S^{n-1},\ \zeta\to \nu_{V,E}(\zeta,x)\cr
&{\rm is\ a\ }C^1{\rm\ orientation\ preserving\ diffeomorphism\ from\ }\pa D {\rm\ onto\ }\S^{n-1}
}\leqno(3.10{\rm b})
$$ 
for $x\in D$ (where we choose the canonical orientation of $\S^{n-1}$ and the orientation of $\pa D$ given by the canonical orientation of $\R^n$ and 
the unit outward normal vector),
$$
\leqalignno{
\nu_{V,E}(\zeta,x)=&{k_0(E,x,\zeta)\over|k_0(E,x,\zeta)|}\cr
=&-{k(E,\zeta,x)\over|k(E,\zeta,x)|},&(3.10{\rm c})
}
$$
for $(\zeta,x)\in \pa D\times D.$ 
Note that from (2.3), (3.9) and (3.10c) one obtains
$$
({\pa l_{V,E}\over \pa x_1}(\zeta,x),..,{\pa l_{V,E}\over \pa x_n}(\zeta,x))={k(E,\zeta,x)\over \sqrt{1-{k(E,\zeta,x)^2\over
c^2}}},\leqno(3.11)
$$
for $(\zeta,x)\in \pa D\times D.$

\vskip 2mm

\noindent 3.5  {\it Determination of an isotropic Riemannian metric.}
We consider the following geometrical problem :
$$
\leqalignno{
&{\rm at\ fixed\ energy\ }E>E(\|V\|_{C^2},D),{\rm\ does\ }l_{V,E}(\zeta,x),{\rm\ given\ for\ all\ }(\zeta,x)\in \pa D\times \pa D,\cr
&{\rm determine\ uniquely\ }r_{V,E}{\rm\ on\ }\bar D\ ?
}
$$
Muhometov-Romanov [MR], Beylkin [B] and Bernstein-Gerver [BG] study the question of determining an isotropic Riemannian metric from its hodograph.
Results in [B] and [BG] are obtained with smoothness conditions that are too strong so that one could apply these results to our problem.
Therefore, for sake of consistency, we give results (Lemma 3.2 and Theorem 3.1) that already appear with stronger smoothness conditions in [B] and [BG].

We denote by $\omega_{0,V}$ the $n-1$ differential form on $\pa D\times D$ obtained in the following manner:

- for $x\in D$, let $\omega_{V,x}$ be the pull-back of $\omega_0$ by $\nu_{V,E,x}$ where $\omega_0$ denotes the canonical orientation form on
$\S^{n-1}$ (i.e. $\omega_0(\zeta)(v_1,..,v_{n-1})={\rm det}(\zeta,v_1,..,v_{n-1}),$ for $\zeta\in \S^{n-1}$ and $v_1,..,v_{n-1}\in T_\zeta\S^{n-1}$),

- for $(\zeta,x)\in \pa D\times D$ and for any $(v_1,..,v_{n-1})\in T_{(\zeta,x)}(\pa D\times D),$  
$$\omega_{0,V}(\zeta,x)(v_1,..,v_{n-1})=\omega_{V,x}(\zeta)(\sigma'_{(\zeta,x)}(v_1),..,\sigma'_{(\zeta,x)}(v_{n-1})),$$
where $\sigma :\pa D\times D\to \pa D,$
$(\zeta',x')\mapsto \zeta',$ and $\sigma'_{(\zeta,x)}$ denotes the derivative (linear part) of $\sigma$ at $(\zeta,x)$.

From smoothness of $\nu_{V,E},$ $\sigma$ and $\omega_0$, it follows that $\omega_{0,V}$ is a continuous $n-1$ form on $\pa D\times D.$

Now let $\lambda\in \R^+$ and $V_1, V_2\in C^2(\bar D,\R)$ such that $\max(\|V_1\|_{C^2},\|V_2\|_{C^2})\le \lambda.$
Let $E>E(\lambda,D).$

Consider the differential forms $\Phi_0$ on 
$(\pa D\times \pa D)\backslash{\bar G}$ and $\Phi_1$ on $(\pa D\times \bar D)\backslash{\bar G}$ defined by
$$
\leqalignno{
\Phi_0(\zeta,x)=&-(-1)^{n(n+1)\over 2}d_x(l_{V_2,E}-l_{V_1,E})(\zeta,x)\wedge d_{\zeta}(l_{V_2,E}-l_{V_1,E})(\zeta,x)\cr
&\wedge \sum_{p+q=n-2}(dd_\zeta l_{V_1,E}(\zeta,x))^p\wedge(dd_\zeta l_{V_2,E}(\zeta,x))^q,&(3.12)
}
$$
for $(\zeta,x)\in (\pa D\times \pa D)\backslash{\bar G},$ where $d=d_\zeta+d_x,$
$$
\leqalignno{
\Phi_1(\zeta,x)=&-(-1)^{n(n-1)\over 2}\left[d_x l_{V_1,E}(\zeta,x)\wedge (dd_\zeta l_{V_1,E}(\zeta,x))^{n-1}\right.
\cr
&+d_x l_{V_2,E}(\zeta,x)\wedge (dd_\zeta l_{V_2,E}(\zeta,x))^{n-1}
-d_x l_{V_1,E}(\zeta,x)\wedge (dd_\zeta l_{V_2,E}(\zeta,x))^{n-1}\cr
&\left.-d_x l_{V_2,E}(\zeta,x)\wedge (dd_\zeta l_{V_1,E}(\zeta,x))^{n-1}
\right],&(3.13)
}
$$
for $(\zeta,x)\in (\pa D\times \bar D)\backslash{\bar G},$ where $d=d_\zeta+d_x.$

From (3.6), (3.7) and (3.8), it follows that $\Phi_0$ is continuous on  $(\pa D\times \pa D)\backslash{\bar G}$ and integrable on $\pa D\times \pa D$
and $\Phi_1$ is continuous on 
$(\pa D\times \bar D)\backslash{\bar G}$ and integrable on $\pa D\times \bar{D}$.

\vskip 2mm

{\bf Lemma 3.2.} {\it Let $\lambda\in \R^+$ and $E>E(\lambda,D).$ Let $V_1,V_2\in C^2(\bar D,\R)$ such that $\max(\|V_1\|_{C^2},\|V_2\|_{C^2})\le
\lambda.$}
{\it The following equalities are valid:
$$
\leqalignno{
\int_{\pa D\times \pa D} \Phi_0=&\int_{\pa D\times \bar{D}}\Phi_1;&(3.14)\cr
{1\over (n-1)!}\Phi_1(\zeta,x)=&\left(r_{V_1,E}(x)^n\omega_{0,V_1}(\zeta,x)
+r_{V_2,E}(x)^n\omega_{0,V_2}(\zeta,x)\right.\cr
&-\nabla_xl_{V_1,E}(\zeta,x)\nabla_xl_{V_2,E}(\zeta,x)&(3.15)\cr
&\left.\times\left(r_{V_1,E}(x)^{n-2}\omega_{0,V_1}(\zeta,x)
+r_{V_2,E}(x)^{n-2}\omega_{0,V_2}(\zeta,x)\right)
\right)\cr
&\wedge dx_1\wedge..\wedge dx_n,
}
$$
for $(\zeta,x)\in \pa D\times D,$  where $\nabla_xl_{V_i,E}(\zeta,x)=({\pa l_{V_i,E}\over \pa x_1}(\zeta,x),..,{\pa l_{V_i,E}\over \pa x_n}(\zeta,x))$ for 
$(\zeta,x)\in \pa D\times D$ and $i=1,2.$
}

\vskip 2mm
Equality (3.14) follows from regularization and Stokes' formula.
Using Lemma 3.2, we obtain the following Theorem of uniqueness and stability. 

\vskip 2mm

{\bf Theorem 3.1.} {\it Let $\lambda\in \R^+$ and $E>E(\lambda,D).$ Let $V_1,V_2\in C^2(\bar D,\R)$ such that $\max({\|V_1\|_{C^2},\|V_2\|_{C^2}})\le
\lambda.$}
{\it The following estimate is valid:
$$
\leqalignno{
\int_D\left(r_{V_1,E}(x)-r_{V_2,E}(x)\right)&\left(r_{V_1,E}(x)^{n-1}-r_{V_2,E}(x)^{n-1}\right)dx\le\cr
&{\Gamma({n\over 2})\over 2\pi^{n\over 2}(n-1)!}
\int_{\pa D\times \pa D}\Phi_0.&(3.16)
}
$$
}
\vskip 2mm

Note that for $V_1, V_2\in C^3$ and $\pa D \in C^\infty$ Theorem 3.1 follows directly from the stability estimate of [B] and [BG] for the problem of determining an isotropic
Riemannian metric from its hodographs. Similar Remarks are also valid for Lemma 3.2.
For $V_1,V_2\in C^2$ and $\pa D\in C^2$ an estimate similar to (3.16) follows directly from a stability estimate of [MR] for the problem of determining an isotropic Riemannian metric
from its hodographs.
As we have followed Gerver-Nadirashvili's framework [GN], we have chosen to extend the related results obtained namely in [BG] and [B] to the case of less smooth
metrics.

\vskip 1cm

\noindent{\sect References}
\vskip 4mm

{\parindent=-1.2cm 
\leftskip=-\parindent 

\leavevmode \hbox to 1.2cm {[A]\hfill}V. I. Arnold, {\it Mathematical Methods of Classical Mechanics}, Springer Verlag New York
Heidelberg Berlin, 1978.

\leavevmode \hbox to 1.2cm {[BG]\hfill}I. N. Bernstein, M. L. Gerver, A condition for distinguishing metrics from hodographs, {\it Comput.
Seismology} {\bf 13}, 50-73 (1980) (Russian).  

\leavevmode \hbox to 1.2cm {[B]\hfill}G. Beylkin, Stability and uniqueness of the solution of the inverse kinematic problem of seismology in higher
dimensions, {\it Zap. Nauchn. Sem. Leningrad. Otdel. Mat. Inst. Steklov. (LOMI)} {\bf 84}, 3-6 (1979) (Russian). English transl.: {\it J. Soviet
Math.} {\bf 21}, 251-254 (1983).  

\leavevmode \hbox to 1.2cm {[E]\hfill}A. Einstein, \" Uber das Relativit\" atsprinzip und die aus demselben gezogenen Folgerungen, 
{\it Jahrbuch der Radioaktivit\" at und Elektronik} {\bf 4},  411-462 (1907).

\leavevmode \hbox to 1.2cm {[GN]\hfill}M.L. Gerver, N. S. Nadirashvili, Inverse problem of mechanics at high energies, 
{\it Comput. Seismology} {\bf 15}, 118-125 (1983) (Russian).

\leavevmode \hbox to 1.2cm {[I]\hfill}H. Isozaki, Inverse scattering theory for Dirac operators, {\it Ann. Inst. H. Poincar\'e
Phys. Th\'eor.}  {\bf 66}:2, 237-270  (1997). 

\leavevmode \hbox to 1.2cm {[J1]\hfill}A. Jollivet, On inverse scattering for the multidimensional relativistic Newton equation at 
high energies, {\it J. Math. Phys. } (to appear).

\leavevmode \hbox to 1.2cm {[J2]\hfill}A. Jollivet, On inverse scattering in electromagnetic field in classical relativistic mechanics at high
energies, 2005 preprint, /math-ph/0506008.

\leavevmode \hbox to 1.2cm {[LL]\hfill}L.D. Landau, E.M. Lifschitz, {\it The Classical Theory of Fields}, Pergamon Press New York, 
1971.

\leavevmode \hbox to 1.2cm {[MR]\hfill}R. G. Muhometov, V.G. Romanov, On the problem of determining an isotropic Riemannian metric in 
$n$-dimensional space, {\it Dokl. Akad. Nauk SSSR} {\bf 243}:1, 41-44 (1978) (Russian). English transl.: {\it Soviet math. Dokl.} {\bf 19}, 1330-1333
(1978). 

\leavevmode \hbox to 1.2cm {[NSU]\hfill}A. Nachman, J. Sylvester, G. Uhlmann, An $n$-dimensional Borg-Levinson theorem, {\it Comm. Math. 
Phys.}  {\bf 115}:4, 595-605 (1988).

\leavevmode \hbox to 1.2cm {[N1]\hfill}R.G. Novikov, A multidimensional inverse spectral problem for the equation
$-\Delta\psi+(V(x)-Eu(x))\psi=0,$ {\it Funktsional. Anal. i Prilozhen} {\bf 22}:4, 11-22, 96 (1988). English transl.: {\it Funct. Anal.
Appl.} {\bf 22}:4, 263-272 (1988).  

\leavevmode \hbox to 1.2cm {[N2]\hfill}R.G. Novikov, Small angle scattering and X-ray transform
in classical mechanics, {\it Ark. Mat.} {\bf 37},  141-169 (1999).

\leavevmode \hbox to 1.2cm {[N3]\hfill}R.G. Novikov, The $\overline\partial$-approach to approximate inverse scattering at fixed energy 
in three dimensions, {\it IMRP Int. Math. Res. Pap.}  {\bf 6}, 287-349 (2005). 

\leavevmode \hbox to 1.2cm {[SU]\hfill}P. Stefanov, G. Uhlmann, Boundary rigidity and stability for generic simple metrics, 
{\it J. Am. Math. Soc.} {\bf 18}: 4, 975-1003 (2005).

\vskip 8mm
\noindent A. Jollivet

\noindent Laboratoire de Math\'ematiques Jean Leray (UMR 6629)

\noindent Universit\'e de Nantes 

\noindent F-44322, Nantes cedex 03, BP 92208,  France

\noindent e-mail: jollivet@math.univ-nantes.fr

\end